\documentclass[11pt]{article}
\usepackage{moriond,epsfig}

\def\be{\begin{equation}}
\def\ee{\end{equation}}
\def\bea{\begin{eqnarray}}
\def\eea{\end{eqnarray}}

\def\la{\,\rlap{\raise 0.5ex\hbox{$<$}}{\lower 1.0ex\hbox{$\sim$}}\,}
\def\ga{\,\rlap{\raise 0.5ex\hbox{$>$}}{\lower 1.0ex\hbox{$\sim$}}\,}

\begin{document}
\vspace*{4cm}
\title{Hard X-ray and Gamma-ray Emission Induced by Ultra-High Energy Protons
in Cluster Accretion Shocks and their Detectability}

\author{S. INOUE$^{1,2}$, F. A. AHARONIAN$^{1}$, N. SUGIYAMA$^{2}$}

\address{$^1$Max-Planck-Institut f\"ur Kernphysik, Postfach 103980, 69029 Heidelberg, Germany\\
$^2$National Astronomical Observatory, 2-21-1 Osawa, Mitaka, Tokyo 181-8588, Japan}

\maketitle

\abstracts{
Protons can be plausibly accelerated to $\sim 10^{18}$-$10^{19}$ eV
in strong accretion shocks surrounding massive clusters of galaxies.
Such protons efficiently produce very high energy pairs through interactions with the CMB,
which then radiate synchrotron and inverse Compton emission with hard spectra,
peaking respectively in hard X-rays and TeV gamma-rays.
We discuss the prospects for detecting this emission from nearby clusters
with current and future facilities such as HESS and NeXT,
which should provide a clear signature of ultra-high energy proton acceleration
and an important probe of magnetic fields in the outermost regions of clusters.
}

\section{Introduction}
All sufficiently massive clusters of galaxies should be surrounded by strong accretion shocks,
which can be interesting sites of high energy particle acceleration and nonthermal emission
on account of their high Mach numbers~\cite{ryu}.
Potential nonthermal emission processes that have been discussed for clusters
include synchrotron and inverse Compton (IC) emission from primary electrons
(i.e. those directly accelerated at shocks),
as well as $\pi^0$ decay emission and emission from secondary $\pi^\pm$ decay pairs
triggered by p-p interactions between nonthermal protons and the intracluster gas~\cite{min}.
The observational evidence for such kinds of emission is currently ambiguous:
known types of diffuse radio emission~\cite{fer} cannot be easily explained by such processes;
reports of nonthermal extreme UV and hard X-ray excess emission~\cite{bow} are still controversial;
and there is no clear case yet for gamma-ray emission from any cluster~\cite{reio}.
Cluster accretion shocks have also been proposed as
candidate acceleration sites of ultra-high energy (UHE) cosmic rays;
however, realistic evaluations of the maximum proton energy fall short of $10^{20}$ eV
by 1-2 orders of magnitude~\cite{krb}.
Here we point out a hitherto unstudied emission process:
synchrotron and IC emission from secondary $e^\pm$ pairs
produced in p-$\gamma$ interactions between $10^{18}$-$10^{19}$ eV
protons accelerated in cluster accretion shocks and the cosmic microwave background (CMB)
(see refs.~\cite{aha,rgd} for a related process).
More details can be found in a forthcoming paper~\cite{ias}.

\section{Model and Results}

We fiducially consider a Coma-like cluster
of mass $M=2 \times 10^{15} M_\odot$ located at distance $D=100$ Mpc,
and assume spherically symmetric, steady state conditions.
The accretion shock occurs at the virial radius $R_s \simeq 3.2$ Mpc,
its velocity relative to the upstream gas is $V_s=(4/3)(GM/R_s)^{1/2} \simeq 2200$ km/s,
and its uncertain magnetic field $B_s$ is taken to be a parameter in the range $0.1$-$1 \mu$G.
The maximum energy of shock accelerated protons is determined by
the balance between the acceleration timescale 
and either the energy loss timescale due to photopair and photopion interactions with the CMB,
or the shock lifetime $t_s$, here taken to be the dynamical time $(4/3)R_s/V_s \simeq 1.9$ Gyr.
This gives $E_{\max} \simeq 4.5 \times 10^{18} {\rm eV}
\ \eta^{-1} (B_s/ 1 \mu{\rm G}) (V_s / 2200 {\rm km/s})^2 (t_s/ 1.9 {\rm Gyr})$,
where $\eta$ is the mean free path for scattering by magnetic irregularities
normalized by the proton gyroradius.
We assume $\eta=1$, as inferred observationally for supernova remnant shocks.
The protons are injected into a spherical volume of radius $R_s$
with energy distribution $\propto E_p^{-2} \exp (-E_p/E_{\max})$,
for duration $t_{inj}=t_s$, and at energy injection rate $L_p=0.1 L_{acc}$, where
$L_{acc} \simeq 2.9 \times 10^{46} {\rm erg \ s^{-1}} (M / 2 \times 10^{15} M_\odot)^{5/3}$
is the gas kinetic energy flux through strong accretion shocks.
The gas density around the accretion shock is taken to be $n=10^{-6} {\rm cm^{-3}}$.
Following ref.~\cite{aha},
we compute proton interactions and associated secondary particle production and emission,
including p-$\gamma$ pair and pion production, p-p gamma-ray and pair production,
synchrotron and IC radiation from all pairs,
as well as intergalactic gamma-ray absorption due to the infrared background (IRB).

\begin{figure}
\psfig{figure=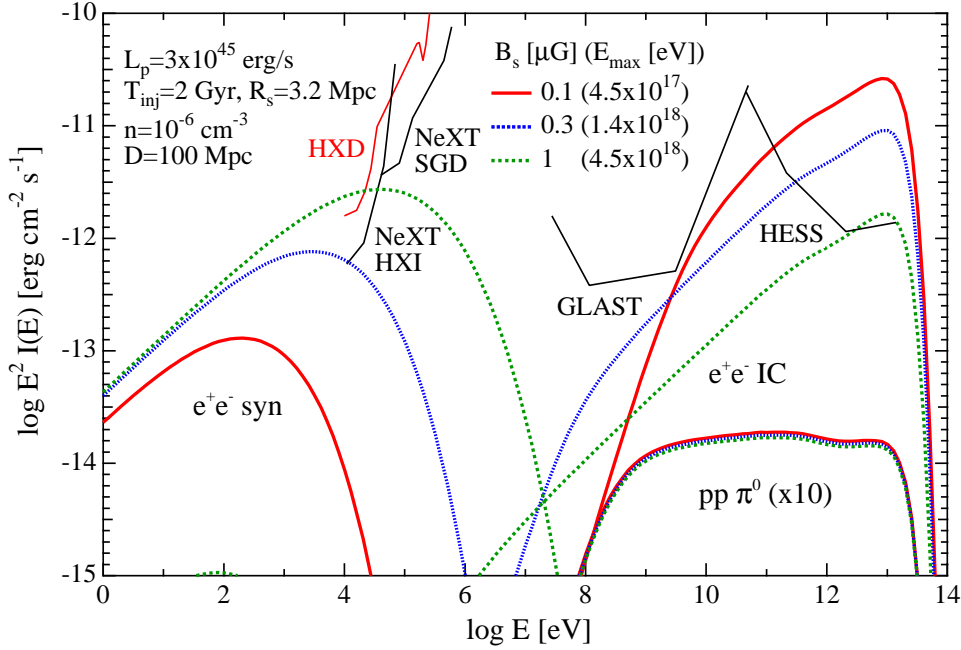,height=3.5in}
\caption{Fiducial model spectra, with the p-p $\pi^0$ component being multiplied by 10.
Overlayed are estimated sensitivities for sources uniformly extended by $1^\circ$,
for HESS, GLAST, Astro-E2/HXD and NeXT.}
\label{fig:flux}
\end{figure}

Fig.\ref{fig:flux} shows the fiducial model spectra for $B_s=0.1$, $0.3$ and $1 \mu$G,
which imply $E_{\max} \sim 10^{18}$-$10^{19}$ eV.
The pairs are injected at energies $\sim (2m_e/m_p) E_p \sim 10^{15}$-$10^{16}$ eV and rapidly cool,
emitting hard spectra with photon indices $\Gamma \sim 1.5$.
Spectral peaks occur at $\sim 1$-$100$ keV for synchrotron and $\sim 10$-$100$ TeV for IC,
with a further cutoff at $\ga 10$ TeV from intergalactic absorption by the IRB.
Note the large influence of $B_s$,
which affects both the synchrotron cooling rate and the amount of pair injection.
The relative luminosities together with the peak frequency locations should allow
$B_s$ to be reliably determined from observations.
The emitted power can reach $\sim 10^{44} {\rm erg \ s^{-1}}$,
indicating the high radiative efficiency of this process induced by protons.
This p-$\gamma$ pair (hereafter PGP) emission
can dominate the hard X-ray and TeV gamma-ray bands
over other components from primary electrons or p-p interactions,
although the latter may still prevail in the GeV range.
The spatial distribution of PGP emission should be similar to primary IC~\cite{min},
tracking the projected ring-like morphology of the accretion shock.

\section{Detectability}

To assess the detectability of PGP emission,
we note that even though the diameter of the accretion shock can be as large as
$3.6^\circ (R_s/3.2 {\rm Mpc}) (D/100 {\rm Mpc})^{-1}$ in the sky,
the expected ring-like distribution is highly nonuniform;
we thus choose to compare model fluxes with instrumental sensitivities for
sources uniformly extended by $1^\circ$.
Plotted in Fig.\ref{fig:flux} are rough estimates of such sensitivities
for various observational facilities discussed below.

Most promising is the PGP IC component in TeV gamma-rays,
where the HESS\footnote{\sf http://www.mpi-hd.mpg.de/hfm/HESS/HESS.html} array of Cerenkov telescopes
can achieve point source sensitivities
$\sim 10^{-13} {\rm erg \ cm^{-2} \ s^{-1}}$ at 1 TeV in $\sim$100 hr exposure time.
With its $0.1^\circ$ angular resolution, the $1^\circ$ extended source sensitivity 
is roughly $10^{-12} {\rm erg \ cm^{-2} \ s^{-1}}$,
so that even our least luminous case of $B_s=1\mu$G may be detectable,
not to mention the more luminous, lower $B_s$ cases.
Furthermore, its $5^\circ$ field of view may allow clear imaging
of the annular morphology with a single pointing.
In contrast, detecting the weaker PGP IC emission at GeV energies may be difficult
even for the GLAST\footnote{\sf http://www-glast.stanford.edu/} mission;
moreover, it could likely be that other emission components such as primary IC
show up above the PGP emission in this band~\cite{min,gb}.
The PGP synchrotron component should be interesting for future facilities such as NeXT~\cite{tak},
capable of imaging hard X-rays in the 8-80 keV range at $\sim$ 30" resolution.
The expected sensitivity for a $1^\circ$ source is
$\sim 10^{-12} {\rm erg \ cm^{-2} \ s^{-1}}$ at $30$ keV in a 100 ks exposure,
so the cases of $B_s \ga 0.3 \mu$G should be clearly detectable,
and the accretion shock morphology may be studied in fine detail.
The non-imaging HXD instrument on the soon-to-be launched Astro-E2~\cite{tak},
with a sensitivity of $\la 3 \times 10^{-12} {\rm erg \ cm^{-2} \ s^{-1}}$ at 20 keV,
as well as instruments onboard INTEGRAL
may also have chances of detection if $B_s \ga 1\mu$G.

\begin{figure}
\psfig{figure=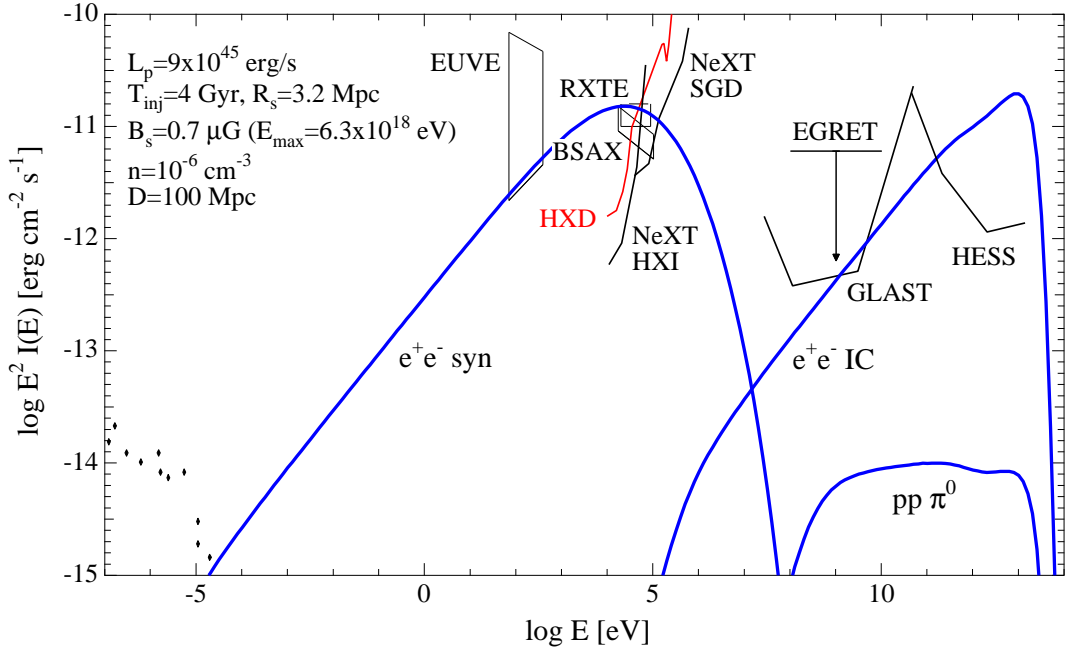,height=3.5in}
\caption{Model spectra as discussed in the text
compared with observational data for the Coma cluster.
Also shown are $1^\circ$ extended source sensitivities as in Fig.\ref{fig:flux}.}
\label{fig:coma}
\end{figure}

Finally, we consider whether the model can explain
the reported excess UV and hard X-ray emission for the Coma cluster.
In Fig.\ref{fig:coma}, multifrequency data~\cite{reia}
are compared with a model with $B_s=0.7 \mu$G, $L_p=9 \times 10^{45} {\rm erg/s}$ and $T_{inj}=4$ Gyr,
otherwise being the same as the fiducial model.
Despite the apparent consistency,
we recall that the UV and hard X-ray data correspond only to the inner $\la 1^\circ$ region,
whereas the model emission should be extended out to radius $\simeq 1.8^\circ$.
To explain the observations with just the fraction of emission projected onto the core,
more extreme parameters for $L_p$ and/or $t_{dyn}$ may be necessary.
In any case, further observations at these as well as gamma-ray energies should provide conclusive answers.
Note that the radio emission cannot be accounted for by PGP emission and requires a different origin.

\section{Implications}

Successful observations of the PGP emission
entails a number of important implications, by providing:
1) a clear signature of UHE protons in cluster accretion shocks,
and hence a test of particle acceleration theory on the largest scales;
2) a sensitive probe of magnetic fields in the outermost regions of clusters, which is crucial for
understanding the origin of intergalactic and intracluster magnetic fields,
propagation of UHE cosmic rays, etc.;
3) a potential probe of the accretion shock itself,
which still lacks firm observational evidence
despite being a robustly predicted cosmological phenomenon;
4) a useful tool for gamma-ray absorption studies of the IRB,
owing to the hard, steady TeV spectra, as opposed to the highly variable spectra of blazars.

\end{document}